\documentclass{ws-rv9x6}
\usepackage{subfigure}     
\usepackage{ws-rv-van}     
\makeindex
\newcommand{\be}{\begin{equation}}
\newcommand{\ee}{\end{equation}}
\def\Tr{{\rm Tr}}

\begin{document}

\chapter[Thermal signatures of pairing correlations in nuclei and nano-scale metallic grains]{Thermal signatures of pairing correlations in\\ nuclei and nano-scale metallic grains}\label{ra_ch1}

\author[Y. Alhassid]{Y. Alhassid}

\address{Center for Theoretical Physics, Sloane Physics Laboratory \\Yale University, New Haven, Connecticut 06520, USA \\
yoram.alhassid@yale.edu}

\begin{abstract}
Atomic nuclei and nano-scale metallic grains are in the crossover regime of pairing correlations between the bulk limit, where the Bardeen-Cooper-Schrieffer (BCS) theory of superconductivity is valid, and the fluctuation-dominated regime, where BCS theory breaks down. In this fluctuation-dominated regime, the pairing gap is comparable to or smaller than the single-particle mean level spacing. We discuss thermal signatures of pairing correlations in nuclei and ultra-small metallic grains that survive despite the large fluctuations of the pairing field.
\end{abstract}

\body

\section{Introduction}\label{intro}

Pairing correlations lead to superconductivity in bulk metals. Effects of the pairing correlations in nuclei, such as a gap in the excitation spectrum of even-even nuclei, are well documented. Superconductivity was explained by the Bardeen-Cooper-Schrieffer (BCS) theory~\cite{BCS}. Following its introduction in electronic systems, the BCS approximation was applied to nuclei by Bohr, Mottelson and Pines~\cite{BMP58} and by Belayev~\cite{Belayev59}.

Single-electron tunneling spectroscopy experiments in ultra-small metallic grains connected to external leads probed the discrete spectra of these grains~\cite{nano-experiments}. A gap was identified in the excitation spectra of larger grains with an even number of electrons. This led to extensive studies of pairing correlations in nano-scale metallic grains~\cite{vondelft01}. Recent technical advances are providing better experimental control over the size and shape of these grains~\cite{Kuemmeth2008}.

BCS theory is a mean-field theory valid in the limit where the pairing gap $\Delta$ is much larger than the single-particle mean level spacing $\delta$. However, in a finite-size system, fluctuations of the order parameter around its mean-field solution can be important. As the linear size of a metallic grain decreases, its single-particle mean level spacing $\delta$ increases. Anderson argued~\cite{Anderson59} that the smallest size at which a metallic grain remains a superconductor corresponds to $\delta \sim \Delta$.  In the smallest metallic grains studied in Ref.~\citeonline{nano-experiments}, the excitation spectrum of the even particle-number grain did not exhibit a noticeable gap when compared to the excitation spectrum of the grain with an odd particle number. These grains belong to the fluctuation-dominated regime $\Delta \lesssim \delta$, in which fluctuations of the pairing field become important and BCS theory breaks down. It was proposed that signatures of pairing correlations in this regime can still be identified through the particle-number parity dependence of thermodynamic observables~\cite{DiLorenzo2000,Falci2000}.

In a nucleus, the gap is typically of the order of the single-particle mean level spacing or somewhat larger. We therefore expect fluctuations in the pairing field to also be important in nuclei.

Here we discuss thermal signatures of pairing correlations in both nuclei and ultra-small metallic grains. We use methods that go beyond the mean-field or BCS approximations and take into account both thermal and quantal fluctuations.  These methods are briefly described in Sec.~\ref{methods}. Applications to nuclei are presented in Sec.~\ref{nuclei}. In Sec.~\ref{Hamiltonian} we discuss the effective configuration-interaction shell model Hamiltonian used in our studies, and in Sec.~\ref{AFMC} we review the auxiliary-field Monte Carlo (AFMC) method used to calculate thermal and statistical nuclear properties in very large model spaces. In Secs.~\ref{heat-capacity}, \ref{spin-distribution} and \ref{moment-of-inertia} we discuss, respectively, signatures of pairing correlations in the heat capacity, the spin distribution of nuclear levels and the thermal moment of inertia. In Sec.~\ref{nanoparticles} we present applications to nano-scale metallic grains. In Sec.~\ref{universal-H} we discuss the universal Hamiltonian describing the effective low-energy Hamiltonian of grains whose single-electron dynamics is chaotic.  Applications of AFMC to calculate signatures of pairing correlations in thermodynamical observables of metallic grains are discussed in Sec.~\ref{nano-AFMC}. The single-particle Hamiltonian of a chaotic grain follows random-matrix theory, and physical observables of the grain undergo mesoscopic fluctuations. The study of these mesoscopic fluctuations requires calculations for a large number of realizations of the single-particle spectrum, and an efficient finite-temperature method is therefore needed. We discuss such a method in Sec.~\ref{finite-T-method}, combining spin and number-parity projections with thermal and small-amplitude quantal fluctuations of the pairing field.  In Sec.~\ref{thermodynamics} we use this method to study the mesoscopic fluctuations of the heat capacity and spin susceptibility of an ultra-small metallic grain. We conclude in Sec.~\ref{conclusion} by comparing thermal signatures of pairing correlations in nuclei with similar signatures in nano-scale metallic grains.

\section{Finite-temperature methods: beyond the mean field}\label{methods}

Correlations beyond the mean-field approximation at finite temperature can be taken into account systematically using the Hubbard-Stratonovich (HS) transformation~\cite{HS}, in which interaction effects are described by including fluctuations around the mean-field solution.

{\em Hubbard-Stratonovich transformation.} The equilibrium density matrix $e^{-\beta H}$ describing a system with an Hamiltonian $H$ at temperature $T=1/\beta$ can be interpreted as the imaginary-time propagator with $\beta$ playing the role of imaginary time. The HS transformation
\begin{equation}\label{HS}
e^{-\beta H} = \int {\cal D}[\sigma] G_\sigma U_\sigma
\end{equation}
expresses this imaginary-time propagator as a coherent superposition of one-body propagators $U_\sigma$ with a Gaussian weight $G_\sigma$. Each $U_\sigma$ describes the imaginary-time propagator of non-interacting particles moving in time-dependent external auxiliary fields $\sigma(\tau)$. We note that the HS decomposition can be done either in the particle-hole channel or in the particle-particle channel.

The HS decomposition~(\ref{HS}) can be used to calculate various thermal observables.  The partition function $Z(T) = {\rm Tr}\, e^{-H/T}$ is found by taking its trace
\be\label{partition}
Z(T) = \int {\cal D}[\sigma] G_\sigma {\rm Tr}\,U_\sigma \;.
\ee

The mean-field approximation is obtained by evaluating the integral in (\ref{partition}) in the saddle-point approximation. In the particle-particle decomposition, this leads to the BCS approximation.

{\em Static path approximation.} To go beyond a mean-field approximation, it is necessary to include fluctuations of the auxiliary fields $\sigma$. At high temperatures, it is sufficient to include static (thermal) fluctuations of the $\sigma$ fields, an approximation known as the static path approximation (SPA)~\cite{Muhlschlegel1972,SPA}. Of particular importance are large-amplitude fluctuations of the order parameter. For example, in the Landau theory of the nuclear shape transitions, static fluctuations in the quadrupole shape parameters were found to be important for understanding the observed temperature and spin dependence of the giant dipole resonance~\cite{GDR}.

{\em Static path approximation plus random phase approximation.} The SPA can be improved by including small-amplitude time-dependent (quantal) fluctuations around each static configuration $\sigma$~\cite{Kerman1981,Lauritzen1990,Puddu1991, Lauritzen1993, Rossignoli1997,Attias1997}. This can be accomplished by expanding $\sigma(\tau) = \sum_r \sigma_r e^{i\omega_r \tau}$ where $\omega_r= 2\pi r /\beta$ ($r$ integer) are the bosonic Matsubara frequencies. For any given static value $\sigma_0$, the integration over $\sigma_r$ ($r \neq 0$) is carried out in the saddle-point approximation, resulting in an $\sigma_0$-dependent random phase approximation (RPA) correction factor. Finally the integration over the static $\sigma_0$ is carried out exactly (i.e., including large-amplitude static fluctuations). This approximation is known as the SPA+RPA. The most important fluctuations originate in those of the order parameters. The approximation breaks down below a  certain critical temperature under which the Gaussian fluctuation in a given $\sigma_r$ (for $r \neq 0$) becomes unstable.

{\em Auxiliary-field Monte Carlo (AFMC).} To account for correlation effects in full, it is necessary to include all  fluctuations -- both thermal and quantal -- of all the auxiliary fields $\sigma$ (including large-amplitude quantal fluctuations). This requires an integration over a very large number of $\sigma$ fields (at all time slices), and in practice can only be done by Monte Carlo methods. Such a quantum Monte Carlo method is generally known as the auxiliary-field Monte Carlo (AFMC) method and has been used in strongly correlated electron systems~\cite{LG92}. In the context of the configuration-interaction shell model the method is known as the shell model Monte Carlo (SMMC) method~\cite{Lang93,Alhassid94,Koonin97,Alhassid01}.

\section{Nuclei}\label{nuclei}

\subsection{Nuclear Hamiltonian}\label{Hamiltonian}

 Here we use the framework of the configuration-interaction shell model approach. The single-particle energies are derived from a central Woods-Saxon potential plus spin-orbit term~\cite{BM69}. Our interaction includes dominant components~\cite{zuker} of effective nuclear interactions: monopole pairing plus multipole-multipole interactions (quadrupole, octupole and hexadecupole)~\cite{NA97}. The latter are obtained by expanding the separable surface-peaked interaction  $v({\bf r}, {\bf r}^\prime) = -\chi (dV/dr)(dV/dr^\prime)\delta(\hat{\bf r} - \hat{\bf r}^\prime)$ ($V$ is the central Woods-Saxon potential) into mulipoles. The coupling constant $\chi$  is determined self-consistently~\cite{ABDK96}
\be
\chi^{-1} = \int_0^\infty dr \; r^2  \left(\frac{dV}{dr}\right)  \left(\frac{d\rho}{dr} \right) \;,
\ee
where $\rho$ is the nuclear density. The quadrupole, octupole and hexadecupole interaction terms are retained and renormalized by factors of $2$, $1.5$ and $1$, respectively.

\subsection{AFMC}\label{AFMC}

For finite-size systems it is often necessary to calculate observables at fixed particle number. The thermal expectation value of an observable $O$ at fixed particle number $A$ is given by
\begin{eqnarray}
\label{eq:obs} \langle O\rangle\equiv \frac{\Tr_A(O e^{-\beta H})}{\Tr_A e^{-\beta H}}=\frac{\int
{\cal D}[\sigma] W_\sigma \Phi_\sigma\langle O\rangle_\sigma}{
\int {\cal D}[\sigma]W_\sigma \Phi_\sigma}\;,
\end{eqnarray}
where $\Tr_A$ denotes a trace at fixed particle number $A$ and we have used the HS transformation (\ref{HS}). Here $W_\sigma=G_\sigma|\Tr_A U_\sigma|$ is a positive-definite weight function,
$\Phi_\sigma=\Tr_A U_\sigma/|\Tr_A U_\sigma|$ is the Monte Carlo ``sign'' and
 $\langle O\rangle_\sigma=\Tr_A (O U_\sigma)/\Tr_A U_\sigma$. The sample-specific quantities $\Tr_A U_\sigma$ and $\langle O\rangle_\sigma$ can be evaluated using matrix algebra in the single-particle space. We denote by ${\bf U}_\sigma$ the $N_{\rm sp}\times N_{\rm sp}$ matrix representing $U_\sigma$ in the single-particle space containing $N_{\rm sp}$ single-particle orbitals. In the grand-canonical ensemble, we then have
\be\label{Tr-U-sigma}
\Tr\, U_\sigma=\det \left(1+{\bf U}_\sigma \right) \;,
\ee
and
\be\label{one-body}
\langle a^\dagger_i a_j\rangle_\sigma = \left(\frac{1}{1+ {\bf U}^{-1}_\sigma}\right)_{ji}\;.
\ee
For a one-body operator $O=\sum_{ij} \langle i |O |j \rangle a^\dagger_i a_j$,  the grand-canonical expectation value $\langle O\rangle_\sigma$ can be calculated using (\ref{one-body}). For a two-body operator, Wick's theorem can be used to express the two-body expectation values in terms of one-body expectation values.

Canonical quantities can be evaluated using particle-number projection with  $\phi_m=2\pi m/N_{\rm sp}$ ($m=1,\ldots,N_{\rm sp}$) as quadrature points to express the canonical trace in terms of grand-canonical traces. For example, the canonical trace of $U_\sigma$ for $A$ particles is given by
\be
\Tr_A U_\sigma= \frac{1}{N_{\rm sp}}\sum_{m=1}^{N_s} e^{-i\phi_m A}
\det\left(1+e^{i\phi_m}{\bf U}_\sigma\right)\;.
\ee

The multi-dimensional integral over the auxiliary fields in (\ref{eq:obs}) is evaluated by Monte Carlo methods. The auxiliary fields are sampled according to the distribution $W_\sigma$. Denoting the samples by $\{\sigma_i\}$, the expectation value in (\ref{eq:obs}) is estimated from
\begin{eqnarray}
\label{eq:obsmc}\langle O\rangle \approx {\sum_i
\Phi_{\sigma_i}\langle O\rangle_{\sigma_i} \over\sum_i
\Phi_{\sigma_i}} \;.
\end{eqnarray}

For a generic interaction, the sign $\Phi_\sigma$ can fluctuate from sample to sample and is in general a phase. At low temperatures, the fluctuations of the sign can become larger than its expectation value. This leads to large statistical errors in thermal observables and the breakdown of the method. When all components of the nuclear interaction discussed in Sec.~\ref{Hamiltonian} are attractive, we have $\Tr U_\sigma >0$ for any $\sigma$, and the interaction has a good Monte Carlo sign in the grand-canonical ensemble. Interactions with small bad-sign components can be treated by the method introduced in Ref.~\citeonline{Alhassid94}.

\subsection{Heat capacity}\label{heat-capacity}

\begin{figure}[h!]
\centerline{\psfig{file=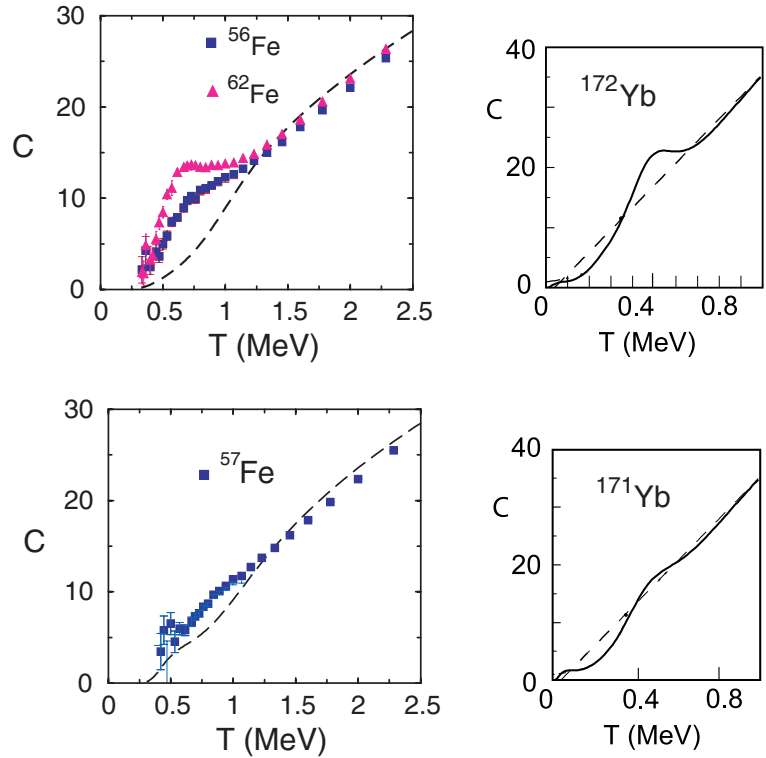,width=10 cm}}
\caption{Heat capacity of even-even (top panels) and odd-even (bottom panels) nuclei. Top: the AFMC heat capacities of $^{56}$Fe and $^{62}$Fe (solid symbols) versus temperature $T$ are compared with the heat capacity of $^{172}$Yb extracted from experiments~\cite{schiller01}. Bottom: As for the top panels but for the odd-even nuclei $^{57}$Fe and  $^{171}$Yb. The dashed lines are obtained in the independent particle model. The shoulder structure observed in the heat capacity of the even-even nuclei is a signature of pairing correlations. Notice that the shoulder structure is enhanced as neutron are added, turning it into a peak for $^{62}$Fe~\cite{la01}.}
\label{hc-nuclei}
\end{figure}

In AFMC, we calculate the thermal energy $E(T)=\langle H\rangle$ as an observable and the heat capacity is determined from $C=dE/dT$. The statistical error in the numerical derivative can be reduced by an order of magnitude by calculating the energy at temperatures $T$ and $T + \delta T$ using the same Monte Carlo sampling and taking into account correlated errors~\cite{la01}.  To obtain the proper behavior of the heat capacity at higher temperatures, we use a method that combines correlated calculations in the truncated model space with independent-particle model calculations in the complete single-particle space~\cite{Alhassid03}.

In Fig.~\ref{hc-nuclei} we show the heat capacities versus temperature for the even-even isotopes $^{56}$Fe and $^{62}$Fe (top left panel) and the odd-even isotope $^{57}$Fe (bottom left panel). In the BCS approximation, the heat capacity is discontinuous at the critical temperature (see the top right panel of Fig.~\ref{nuclei-nanoparticles} where we observe two discontinuities, one for the neutron pairing transition and a second for the proton pairing transition). The AFMC results show significant suppression of the BCS heat capacity because of the large fluctuations in the pairing gap. However, in the even-even nuclei $^{56}$Fe and $^{62}$Fe, there remains a shoulder in the heat capacity despite the large fluctuations. This shoulder structure is also refers to as an $S$-shape heat capacity and was observed experimentally in even-even rare-earth nuclei using the Oslo method~\cite{schiller01} (see, e.g., the heat capacity of $^{172}$Yb in the top right panel of Fig.~\ref{hc-nuclei}). For the odd-even nucleus $^{57}$Fe (bottom left panel of Fig.~\ref{hc-nuclei}), the shoulder structure is suppressed, in qualitative agreement with the experimental result in $^{171}$Yb (bottom right panel).

\subsection{Spin distribution}\label{spin-distribution}

The spin distribution $\rho_J/\rho$ of nuclear energy levels ($\rho$  and $\rho_J$ are, respectively, the total state density and density of levels with spin $J$) versus spin $J$ at a given excitation energy $E_x$ can be calculated exactly in AFMC using a spin projection method~\cite{Alhassid07}. We use the following identity for a scalar operator $X$
\be\label{spin-projection}
\Tr_J  X= \Tr_{M=J}  X - \Tr_{M=J+1} X  \;,
\ee
where $\Tr_J$ denotes trace at a fixed spin $J$, while $\Tr_M$ denotes a trace at a fixed value $M$ of the spin component $J_z$.  The projection $P_M$ on a given value $M$ of $J_z$ is accomplished using the Fourier sum
\be\label{P_J_z}
P_M = \frac{1}{2J_{\rm max}+1}\sum_{m=-J_{\rm max}}^{J_{\rm max}} e^{-i\phi_m M} e^{i\phi_m J_z} \,,
\ee
where $J_{\rm max}$ is the maximal value of the spin in the many-particle model space and  $\phi_m = 2\pi m/(2J_{\rm max}+1)$ are quadrature points.

\begin{figure}[h!]
\centerline{\psfig{file=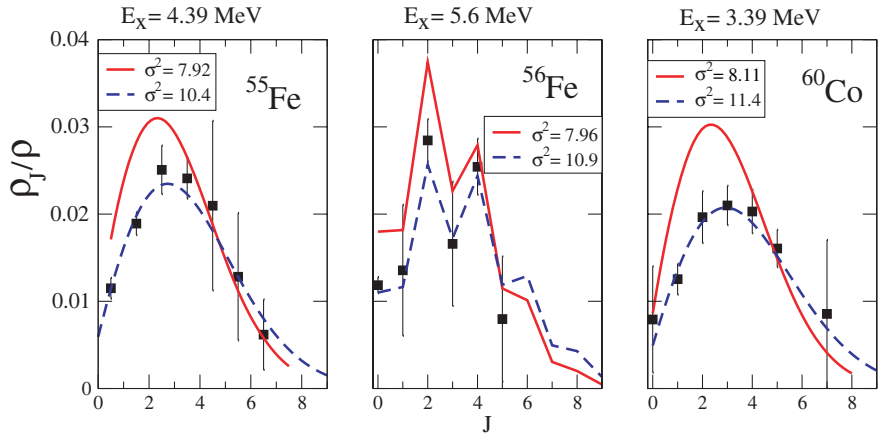,width=12 cm}}
\caption{Spin distributions $\rho_J/\rho$ of energy levels in $^{55}$Fe, $^{56}$Fe and $^{60}$Co. Solid squares are the AFMC results at the excitation energies indicated in the figure. The solid lines are empirical distributions (see text) deduced from global fits to experimentally known low-lying levels~\cite{vonEgidy08}. The dashed lines are the empirical distributions but with the higher $\sigma^2$ values shown in the legends. These higher values are consistent with the higher excitation energies of the AFMC results. From Ref.~\citeonline{vonEgidy08}.}
\label{spin}
\end{figure}

In Fig.~\ref{spin} we show by solid squares the AFMC spin distributions for an odd-even nucleus ($^{55}$Fe), an even-even nucleus ($^{56}$Fe) and an odd-odd nucleus ($^{60}$Co) at excitation energies of $E_x=4.39$,  $5.6$, and $3.39$ MeV, respectively~\cite{Alhassid07}.

\begin{figure}[h]
\centerline{\psfig{file=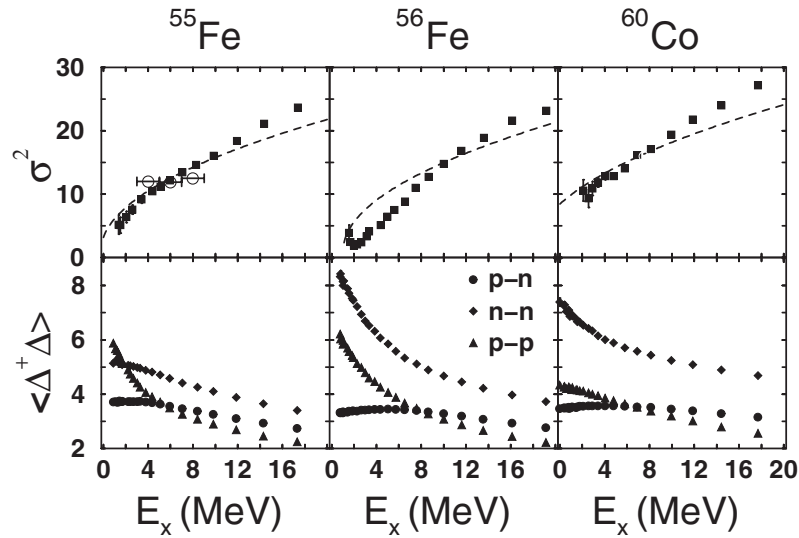,width=10.5 cm}}
\caption{The squared spin-cutoff parameter $\sigma^2$ (top panels) extracted from the AFMC spin distributions (solid squares) and the $J=0$ pair correlations $\langle \Delta^\dagger \Delta\rangle$ (bottom panels) versus excitation energy for $^{55}$Fe, $^{56}$Fe and $^{60}$Co. The dashed lines correspond to $\sigma^2=I T/\hbar^2$ with rigid-body moment of inertia $I$. The open circles in the $\sigma^2$ panel of $^{55}$Fe are experimental data~\cite{gr74}. From Ref.~\citeonline{Alhassid07}.}
\label{sigma-delta}
\end{figure}

We compare our results with the spin-cutoff model obtained through the random coupling of the single-nucleon spins to total spin $\bf J$~\cite{er60}. In this model
\be
  \label{spin-cutoff}
  \frac{\rho_J}{\rho} =
  {(2J+1) \over 2\sqrt{2 \pi} \sigma^3} e^{-{J(J+1) \over 2\sigma^2}}\;,
\ee
where $\sigma$ is the spin-cutoff parameter.  The AFMC spin distribution for odd-even and odd-odd nuclei are well described by the spin-cutoff model (\ref{spin-cutoff}) with a fitted energy-dependent parameter $\sigma=\sigma(E_x)$ (dashed lines in Fig.~\ref{spin}). The energy dependence of $\sigma^2$ (extracted from fits to the AFMC spin distributions) is shown in Fig.~\ref{sigma-delta}.  The solid lines in Fig.~\ref{spin} are the spin-cutoff formula with an empirical value of $\sigma^2=2.61 A^{0.28}$ as determined from global fits to spin distributions of experimentally known low-lying levels~\cite{vonEgidy08}. These distributions agree well with the AFMC distributions (solid squares) once the value of $\sigma^2$ is scaled to larger values shown in the legends to take into account the higher excitation energies of the AFMC results. These empirical distributions with the scaled values of $\sigma^2$ are shown by the dashed lines in Fig.~\ref{spin})

For even-even nuclei (e.g., $^{56}$Fe), the spin-cutoff model works well only at higher excitation energies. As the excitation energy is lowered, an odd-even staggering effect as a function of spin is observed in the AFMC calculations. Such staggering effect was confirmed in the empirical studies of Ref.~\citeonline{vonEgidy08} and was parametrized by multiplying the spin-cutoff formula (\ref{spin-cutoff}) by $1+x$ where $x=0.227\; (-0.277)$ for even (odd) non-zero spin values and $x=1.02$ for $J=0$.  This empirical distribution is shown by the solid line in Fig.~\ref{spin} for $^{56}$Fe. The same distribution, scaled to a larger value of $\sigma^2$ (dashed line) agrees well with the AFMC results.

Fig.~\ref{sigma-delta} shows the energy dependence of $\sigma^2$ as extracted from the AFMC spin distributions (solid squares).  They are compared with the curves $\sigma^2(E_x)$ calculated from $\sigma^2=I T /\hbar^2$ using a rigid-body moment of inertia $I$ (dashed lines). For the odd-even and odd-nuclei we observe general agreement. However, for the even-even nucleus, we find that  $\sigma^2$ is suppressed at low excitation energies (below the pairing transition) when compared to its rigid-body value. The corresponding suppression of the moment of inertia is correlated with the onset of large neutron pair correlations $\langle \Delta^\dagger \Delta\rangle$ (here $\Delta^\dagger = \sum_{a m_a>0} (-1)^{j_a - m_a} a^{\dagger}_{j_a m_a}a^{\dagger}_{j_a -m_a}$ is the $J=0$ pair operator) at low excitation energies (see bottom middle panel of Fig.~\ref{sigma-delta}) and is a signature of pairing correlations.

\subsection{Thermal moment of inertia}\label{moment-of-inertia}

\begin{figure}[t!]
\centerline{\psfig{file=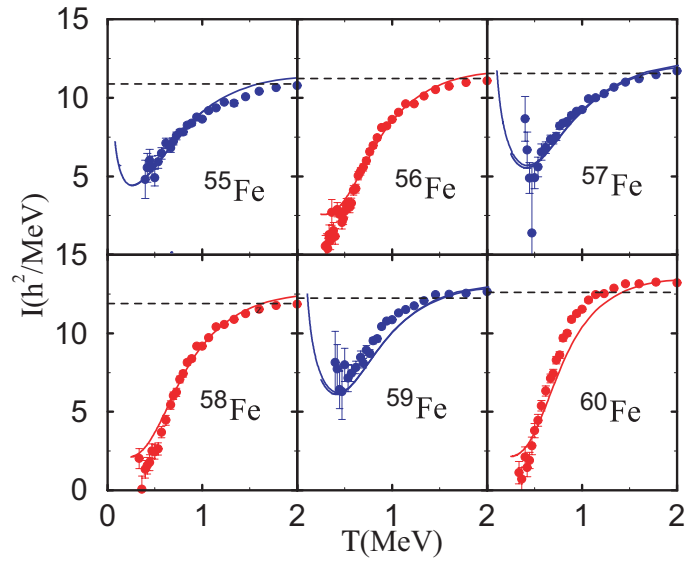,width=9 cm}}
\caption{The thermal moment of inertia of iron isotopes versus temperature $T$. The symbols are the AFMC results and the lines are from a simple model described in the text. The dotted-dashed line are the rigid-body moments of inertia. Note the reentrant behavior of the moment of inertia of the odd-mass isotopes. From Ref.~\citeonline{Alhassid05}.}
\label{inertia}
\end{figure}

The moment of inertia describes the response of the nucleus to rotations.  At finite temperature and for a rotationally-invariant Hamiltonian, it is given by $I=\beta \langle J_z^2\rangle$. In Fig.~\ref{inertia} we show the AFMC results (circles) for $I$ in iron nuclei. These results can be explained by a simple model (lines), in which we consider a monopole pairing Hamiltonian for a deformed nucleus~\cite{Alhassid05}. We then use the SPA together with a number-parity projection to describe the odd-even effects. In even-even nuclei $I$ is suppressed at low temperatures because of pairing correlations, while in odd-even nuclei the suppression is weaker and in the limit $T \to 0$ we observe an enhancement because of the unpaired nucleon.

\section{Nano-scale metallic grains}\label{nanoparticles}

The spectra of nano-scale metallic grains were determined as a function of a Zeeman magnetic field by measuring the non-linear conductance of grains connected to external leads~\cite{nano-experiments}. In larger grains, a pairing gap was observed in the excitation spectrum of a grain with an even number of electrons. However, in smaller grains where the pairing gap $\Delta$ becomes comparable to the mean level spacing $\delta$, it is difficult to resolve such a gap. These grains describe the crossover between the bulk BCS limit and the fluctuation-dominated regime, in which BCS theory is no longer valid and the effects of pairing correlations are much more subtle.

\subsection{Universal Hamiltonian}\label{universal-H}

Here we discuss a metallic grain whose single-electron dynamics is chaotic.  The single-particle energies $\epsilon_i$ and orbital wave functions of such a  grain exhibit mesoscopic fluctuations (from sample to sample or as a function of energy) that follow random-matrix theory (RMT)~\cite{Alhassid2000}. The two-body electron-electron interaction matrix elements, when expressed in the basis of single-particle eigenstates of the one-body Hamiltonian, fluctuate too. We can decompose these interaction matrix elements into an average and fluctuating parts~\cite{Kurland2000,Aleiner2002,Alhassid2005}. The fluctuating part of the interaction matrix elements can be shown to be suppressed by $1/g_T$, where $g_T$ is the Thouless conductance of the grain. Here we consider the limit of large $g_T$, where the fluctuating part of the interaction can be ignored.

The single-particle Hamiltonian together with the average part of the interaction defines the so-called universal Hamiltonian~\cite{Kurland2000,Aleiner2002}. This effective Hamiltonian describes the low-energy physics of a chaotic metallic grains and contains three interaction terms: charging energy, spin exchange and pairing. For a fixed number of electrons, the charging energy is constant and the universal Hamiltonian has the following form
\be\label{universal_hamiltonian}
 H = \sum_{i, \sigma=\uparrow,\downarrow} \epsilon_i c^\dagger_{i\sigma}
c_{i\sigma} - g  P^\dagger P - J_s {\bf S}^2\,,
\ee
where
\be \label{pair-operator}
P^\dagger = \sum_i c^\dagger_{i\uparrow}c^\dagger_{i\downarrow}
\ee
 is the pair operator and  ${\bf S}$ is the total spin of the grain.

Pairing correlations, which favors a superconducting minimal-spin ground state, compete with the exchange interaction, which favors a ferromagnetic spin-polarized state. At zero temperature this competition leads to a coexistence regime in which the ground-state wave function is partly spin-polarized and partly paired~\cite{Ying2006,Schmidt2007}. Here we discuss signatures of this competition in thermodynamic observables (such as heat capacity and spin susceptibility) and their mesoscopic fluctuations.

\subsection{AFMC}\label{nano-AFMC}

In the absence of exchange correlations ($J_s=0$), the universal Hamiltonian (\ref{universal_hamiltonian}) reduces to a BCS-like Hamiltonian. Signatures of the BCS-like interaction in various thermodynamic observables of the grain were studied by a number of methods~\cite{Muhlschlegel1972,DiLorenzo2000, Falci2000, Falci2002, VanHoucke2006, Schechter2001}. Here we discuss an application of the AFMC method~\cite{Alhassid2007} (an attractive BCS-like interaction has a good Monte Carlo sign in the density decomposition of the HS transformation). The band width of the grain is determined by the Debye frequency. However, in practical calculations we truncate the band width to $2N_r+1$ single-particle levels, renormalizing the coupling constant $g$ to keep the BCS gap fixed for the discrete system.  For half filling of the band, the renormalized coupling constant $g_r$ is given by~\cite{Berger1998,Alhassid2007}
\be
\frac{g_r}{\delta} = \frac{1}{{\rm arcsinh}\left(\frac{N_r+1/2}{\Delta/\delta}\right) } \;.
\ee

The heat capacity is calculated as in the nuclear case (see Sec.~\ref{heat-capacity}), while the spin susceptibility $\chi=d M/dB |_{B=0}$ ($M$ is a magnetization for a weak Zeeman field $B$)  is calculated from
\be\label{chi}
\chi= 4 \beta \mu_B^2 \left( \langle S_z^2\rangle - \langle S_z\rangle^2 \right)\;.
\ee
 Here $\mu_B$ is the Bohr magneton and $ S_z$ is the $z$ component of the total spin.

\begin{figure}[t!]
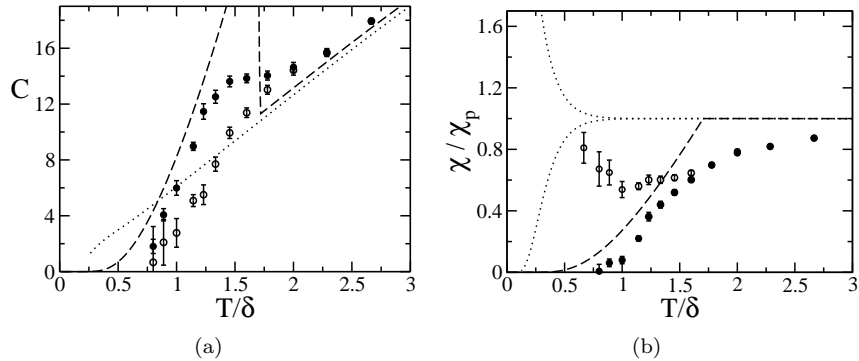

\centerline{
  \subfigure[]
     {\epsfig{figure=hc_3.eps,width=5.4 cm}}
  \hspace*{4pt}
  \subfigure[]
     {\epsfig{figure=sus_3.eps,width=5.5 cm}}
}
\caption{(a) Heat capacity $C$ and (b) spin susceptibility  $\chi/\chi_P$ (measured in units of the Pauli susceptibility $\chi_P=2\mu_B^2/\delta$) of a metallic grain with $\Delta/\delta=3$ and equally spaced single-particle spectrum. Solid (open) circles are the AFMC results for a grain with even (odd) particle number. The dotted lines are the results of the canonical independent-particle model and the dashed lines correspond to the BCS approximation. From Ref.~\citeonline{Alhassid2007}. }\label{hc_sus}
\end{figure}

The AFMC heat capacity (left panel) and spin susceptibility (right panel) are shown in Fig.~\ref{hc_sus} for a grain with $\Delta/\delta=3$ and an equally spaced single-particle spectrum. The solid (open) circles correspond to an grain with and even (odd) number of electrons. The BCS heat capacity [dashed line in (a)] displays a discontinuity at the critical temperature for superconductivity. In the finite grain, the heat capacity is a smooth function of temperature but exhibits an odd-even effect that is a signature of pairing correlations. In the odd grain, the spin susceptibility exhibits a re-entrance effect: it decreases as the temperature is lowered before diverging in the limit $T \to 0$.

\subsection{A finite-temperature method}\label{finite-T-method}

An attractive (ferromagnetic) exchange interaction leads to a sign problem in AFMC, but another quantum Monte Carlo method, suitable for a pairing interaction,  is free of such a sign problem and was used to calculate thermodynamic observables of the grain~\cite{VanHoucke2010}.  The energy eigenvalues of the universal Hamiltonian can also be determined by generalizing Richardson's solution~\cite{richardson} to include the exchange interaction~\cite{Alhassid2003,Schmidt2007}.  However, to study the mesoscopic fluctuations, it is necessary to repeat the calculations for a large number of samples of the single-particle Hamiltonian. Both the quantum Monte Carlo method and Richardson's solution are computationally intensive. We used instead a finite-temperature method~\cite{Nesterov2012}, in which the exchange interaction is treated exactly by spin projection~\cite{Alhassid07}, and the pairing interaction is solved in the SPA+RPA together with a number-parity projection~\cite{Goodman1981,Rossignoli1998,Balian1999,Alhassid05}. The number-parity projected SPA+RPA method was used in Ref.~\citeonline{Falci2002} to study odd-even effects in thermodynamic properties of a metallic grain in the absence of exchange correlations.

\subsubsection{Spin projection}\label{spin-project}

We treat the exchange interaction exactly using the spin projection method of Sec.~\ref{spin-distribution}. The partition function $Z = \Tr\, e^{-\beta \hat H}$ of the universal Hamiltonian at temperature $T=1/\beta$ can be written as
\be\label{spin-proj}
Z = \sum_{S} e^{\beta J_s S(S+1)} \Tr_S e^{-\beta H _{\rm BCS}}\,,
\ee
where $\Tr_S$ is the trace over states with fixed spin $S$, and
$H_{\rm BCS}$ is the BCS-like pairing Hamiltonian
\be\label{BCS_Hamiltonian}
    H_{BCS} = \sum_{i,\sigma=\uparrow,\downarrow} \epsilon_i c^\dagger_{i\sigma}
    c_{i\sigma} - g P^\dagger P\,.
\ee
Using the identity (\ref{spin-projection}) with $S$ replacing $J$ and $X = e^{-\beta H _{\rm BCS}}$, we can rewrite (\ref{spin-proj}) in the form
\be\label{spin-proj-Z}
Z = \sum_S e^{\beta J_s S(S+1)} \left( Z_{M=S}  -
Z_{M=S+1} \right) \;.
\ee
 Here $Z_M = \Tr_M \left( e^{-\beta\hat H _{\rm BCS}}\right)$ with the trace taken at fixed value $M$ of the spin component $S_z$.
The $S_z$ projection is given by a formula similar to (\ref{P_J_z}) but with $S_z$ replacing $J_z$.

The spin susceptibility $\chi$ can be calculated from Eq.~(\ref{chi}) using $\langle S_z^2\rangle = \langle {\bf S}^2\rangle/3$ and $\langle S_z\rangle=0$. We find
\begin{eqnarray}\label{chi_proj}
\chi= \frac{4\beta\mu_B^2}{3 Z} \sum_S S(S+1)(2S+1)
  e^{\beta J_s S(S+1)}\left( Z_{M=S}-Z_{M=S+1} \right)\,.
\end{eqnarray}

\subsubsection{Number-parity projection}\label{number-parity}

We carry out particle-number projection in the saddle-point approximation, where the canonical ensemble is approximated by the grand-canonical ensemble with an average particle number $N$. To describe the odd-even effects of pairing correlations, we use the number-parity projection~\cite{Goodman1981,Rossignoli1998,Balian1999,Alhassid05}
\be
P_\eta= \frac{1}{2}\left(1 + e^{i\pi \hat N}\right)\;,
\ee
with $\eta=1$ ($\eta=-1$) corresponding to projection on an even (odd) number of particles and $\hat N$ is the particle-number operator.

\subsubsection{Hubbard-Stratonovich transformation in the pairing channel}\label{pairing-HS}

The HS decomposition we use in the AFMC applications to nuclei and nano-scale metallic grains  corresponds to a density decomposition, in which the auxiliary fields $\sigma(\tau)$ are densities.  The reason for doing so is that there is no Monte Carlo sign problem in such a decomposition. Here we use a pairing decomposition, in which the auxiliary field is the complex pairing field $\Delta(\tau)$.  In this pairing decomposition, the propagator of the BCS Hamiltonian in the grand-canonical formalism is given by
\be\label{HS-pairing}
 e^{-\beta \left( H_{\rm BCS} - \mu \hat N\right)} = \int {\mathcal{D}}[\Delta,\Delta^*] e^{- \int\limits_0^\beta d\tau |\Delta(\tau)|^2/g} U_\Delta\,.
\ee
Here $U_\Delta = {\cal T}e^{- \int\limits_0^\beta d\tau\, H_{\Delta(\tau)} }$ (${\cal T}$ denotes time ordering) is the propagator for the one-body Hamiltonian
\begin{eqnarray}\label{H_eff(xi_1,xi_2)}
\hat{H}_{\Delta(\tau)}
 =  \sum_i && \left[\left(\epsilon_i - \mu-\frac
g2\right)\left(c^\dagger_{i\downarrow}c_{i\downarrow}+
c^\dagger_{i\uparrow}c_{i\uparrow}\right) \right. \nonumber \\& & -  \left. \Delta(\tau)\, c^\dagger_{i\uparrow} c^\dagger_{i\downarrow} - \Delta^*(\tau) \,c_{i\downarrow}c_{i\uparrow}+
\frac{g}{2}\right]\;.
\end{eqnarray}

\subsubsection{SPA+ RPA}\label{SPA+RPA}

The number-parity and $S_z$-projected partition function $Z_{\eta, M} = \Tr \left[P_\eta P_{M} e^{-\beta \left( H_{\rm BCS} - \mu \hat N\right)}\right]$ can be calculated using the HS transformation (\ref{HS-pairing}) as follows. We expand the pairing field in a Fourier series $\Delta(\tau) = \Delta_0 + \sum_{r \neq 0} \Delta_r e^{i \omega_r \tau}$, where $\omega_r=2\pi r/\beta$ ($r$ integer) are bosonic Matsubara frequencies. For each static fluctuation $\Delta_0$, we perform the integration over $\Delta_r$ ($r \neq 0$) in the saddle-point approximation and then keep the exact integral over $|\Delta_0|$ (the integration over the phase of $\Delta_0$ is trivially done).  The projection on particle number $N$ is performed in a saddle-point approximation, leading to the following expression for the $N$-particle partition function at fixed $\eta$ and $M$~\cite{Nesterov2012}
\begin{eqnarray}\label{Z_N}
 Z_{N,\eta,M}  \approx  \int \limits_0^\infty & &\frac{\beta \,d\,|\Delta_0|^2}{g}  \left(\frac{2\pi}{\beta}\left|\frac{\partial^2 F}{\partial \mu^2}\right|\right)^{-1/2} \nonumber \\
 & &\times \; e^{-(\beta/g)|\Delta_0|^2} \,\, e^{-\beta \mu N}  Z_{\eta, M}(\Delta_0) \,\,C^{\mathrm{RPA}}_{\eta, M}(\Delta_0)\;.
\end{eqnarray}
 Here $Z_{\eta, M}(\Delta_0) $ is the number-parity and $S_z$-projected partition function for a static fluctuation $\Delta_0$
\begin{eqnarray}\label{Z(Dstatic)}
 Z_{\eta, M}(\Delta_0) & = &\left[\prod_i e^{-\beta(\epsilon_i-\mu-E_i)}\right] \left[\sum_{m} \frac{e^{-i\phi_m M}}{2(2S_{\rm max}+1)} \right.\nonumber \\ &\times & \left. \left(\prod_i \left| 1+e^{-\beta E_i + \frac{i\phi_m}{2}}\right|^2 + \eta \prod_i \left| 1- e^{-\beta E_i + \frac{i\phi_m}{2}}\right|^2 \right)\right]\;
\end{eqnarray}
 where $ E_i = \sqrt{\left(\epsilon_i - \mu - \frac g2\right)^2 + |\Delta_0|^2} $ are the quasiparticle energies for the given $\Delta_0$.  The factor $C^{\mathrm{RPA}}_{\eta, M}(\Delta_0)$ is a local RPA correction factor, arising from the small-amplitude time-dependent fluctuations \be\label{C_RPA = sinh/sinh}
C^{\mathrm{RPA}}_{\eta, M} (\Delta_0) = \prod_i
\frac{\Omega_i}{2E_i}\frac{\sinh(\beta
E_i)}{\sinh\left(\frac{\beta \Omega_i}2\right)}\,.
\ee
The local RPA frequencies $\pm \Omega_i$ are the eigenvalues of
the $2N_{\text{sp}}\times 2N_{\text{sp}}$ RPA matrix ($N_{\text{sp}}$ is the number of single-particle orbitals)
\be\label{RPA_matrix}
\left(
\begin{array}{cc}
  2E_i \delta_{ij} - \frac g2f_{i}(\gamma_i\gamma_j+1) & -\frac g2 f_{i}(\gamma_i\gamma_j-1)  \\
  \frac g2 f_{i}(\gamma_i\gamma_j-1)  & \frac g2 f_{i}(\gamma_i\gamma_j+1)- 2E_i \delta_{ij} \\
\end{array}%
\right)
\ee
with $\gamma_i = \left(\epsilon_i - \mu - \frac g2\right)/E_i$ and
\be\label{F_eta_i}
f_{i} = \frac{1}{\beta} \frac{\partial \ln Z_{\eta,M}(\Delta_0)}{\partial E_i}\,.
\ee
$F$ in Eq.~(\ref{Z_N}) is the grand-canonical free energy
\be
 F = \frac{|\Delta_0|^2}{G} +  \sum_i \left[(\epsilon_i -\mu) - 2\beta^{-1}\ln \left(2\cosh\frac{\beta E_i}{2} \right) \right]\;,
\ee
and $\mu$ is the chemical potential determined (for each $\Delta_0$) from the particle number equation $N=-\partial F/\partial \mu$.

The partition function and spin susceptibility for a metallic grain with $N$ electrons and number parity $\eta$ are calculated from Eqs.~(\ref{spin-proj-Z}) and (\ref{chi_proj}), respectively, where $Z_M$ is the partition function $Z_{N,\eta,M}$ in Eq.~(\ref{Z_N}).

The SPA+RPA method breaks down below a critical temperature when a fluctuation $\Delta_r$ for a certain value of  $\Delta_0$  becomes unstable.  A method was recently proposed~\cite{Ribeiro2012} to overcome this problem.

\begin{figure}
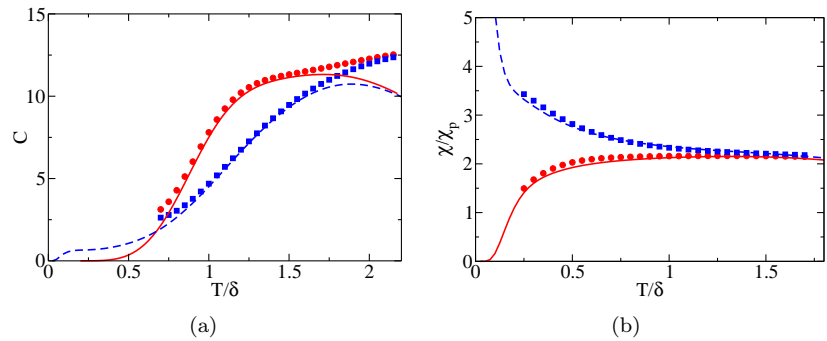

\centerline{
 \subfigure[]
     {\epsfig{figure=hc_rich.eps,width=5.2 cm}}
  \hspace*{5.2 pt}
  \subfigure[]
     {\epsfig{figure=ss_rich.eps,width=5.2 cm}}
}
\caption{The results of the spin and number-parity projected SPA+RPA method (solid circles for an even particle number and solid squares for an odd particle number) are compared with exact results obtained from Richardson's solution (solid lines for an even grain and dashed lines for an odd grain): (a) heat capacity for a grain with $\Delta/\delta=3$ and $J_s/\delta=0.5$; (b) spin susceptibility $\chi/\chi_P$ for a grain with $\Delta/\delta=0.5$ and $J_s/\delta=0.5$. The single-particle spectra correspond to specific RMT realizations. From Ref.~\citeonline{Nesterov2012}.} \label{hc_ss_rich}
\end{figure}

In Fig.~\ref{hc_ss_rich} we compare the number-parity projected SPA+RPA results (symbols) with the exact results obtained from Richardson's solution (lines). We see that this approximation is very good except that it breaks down below a certain critical temperature. However, the odd-even signatures of interest are observed above this temperature. In using Richardson's solution we calculate all the eigenvalues of the universal Hamiltonian below an energy cutoff of $\sim 30 \,\delta$ so the corresponding results in Fig.~\ref{hc_ss_rich} are no longer accurate above $T/\delta \sim 1.5$. The correct heat capacity is then given by the number-parity projected SPA+RPA results.\\

\subsection{Thermal observables: heat capacity and spin susceptibility}\label{thermodynamics}

\begin{figure}
\centerline{\psfig{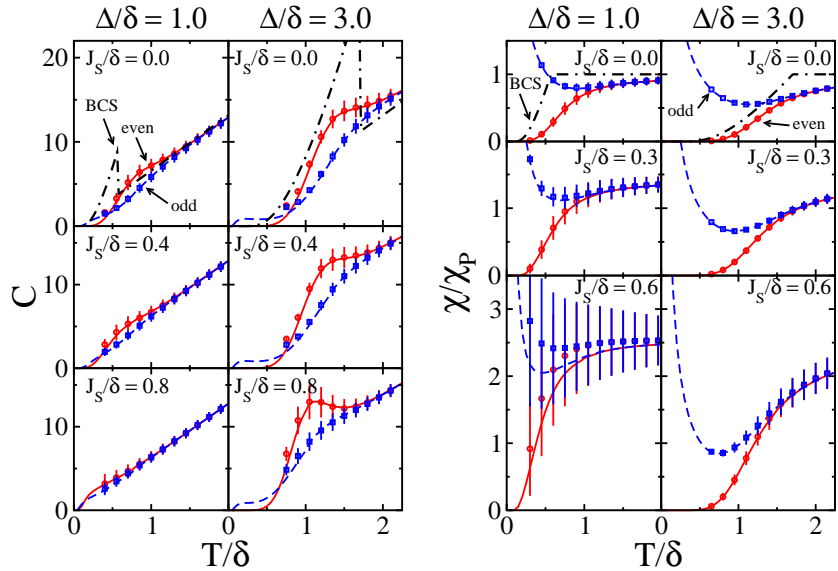}}
\caption{Mesoscopic fluctuations of (a) heat capacity $C$ and (b) spin susceptibility $\chi/\chi_P$ in metallic grains with $\Delta/\delta=1$ (left columns)  and  $\Delta/\delta=3$ (right columns). Results are shown for different value of $J_s/\delta$ and for both even grains (circles) and odd grains (squares).  The symbols and vertical bars describe, respectively, the average values and standard deviations (over the ensemble of single-particle spectra). The lines are the results for an equally spaced single-particle spectrum and the dotted-dashed lines are the grand-canonical BCS results. From Ref.~\citeonline{Nesterov2012}.}
\label{mes_fluct}
\end{figure}

We used the method of Sec.~\ref{finite-T-method} to study the mesoscopic fluctuations of the heat capacity and spin susceptibility of a metallic grain for a large number of realizations of the single-particle RMT spectrum~\cite{Nesterov2012}.  The results are summarized in Fig.~\ref{mes_fluct} for both even and odd grains and for different values of $\Delta/\delta$ and $J_s/\delta$. The symbols are the average values over the ensemble and the vertical bars describe the standard deviations of the corresponding quantities. The lines are the results for an equally spaced single-particle spectrum.

As already discussed in Sec.~\ref{nano-AFMC}, pairing correlations lead to odd-even effects in the heat capacity and spin susceptibility. Here we study how exchange correlations and mesoscopic fluctuations affect these number-parity dependent signatures of pairing correlations.

In general, the exchange interaction shifts the odd-even effects in the heat capacity and spin susceptibility to lower temperatures.  In the fluctuation-dominated regime $\Delta/\delta \lesssim 1$, exchange correlations suppress the odd-even effect in the heat capacity as well as the reentrant behavior of the spin susceptibility for an odd number of electrons. However, for $\Delta >\delta$, exchange correlations enhance the shoulder in the even particle-number heat capacity and can turn it into a peak (see the panel in Fig.~\ref{mes_fluct} with $\Delta/\delta=3$ and $J_s/\delta=0.8$). Similarly for $\Delta >\delta$, exchange correlations enhance the reentrant effect for an odd number of electrons (see the panel with $\Delta/\delta=3$ and $J_s/\delta=0.6$).

In the fluctuation-dominated regime,  $\Delta/\delta \lesssim 1$, the mesoscopic fluctuations of the heat capacity can wash out the odd-even effect for moderate strengths of the exchange interaction. In this regime we also observe large fluctuations of the spin susceptibility as the exchange coupling constant increases.

\section{Conclusion}\label{conclusion}

\begin{figure}
\centerline{\psfig{file=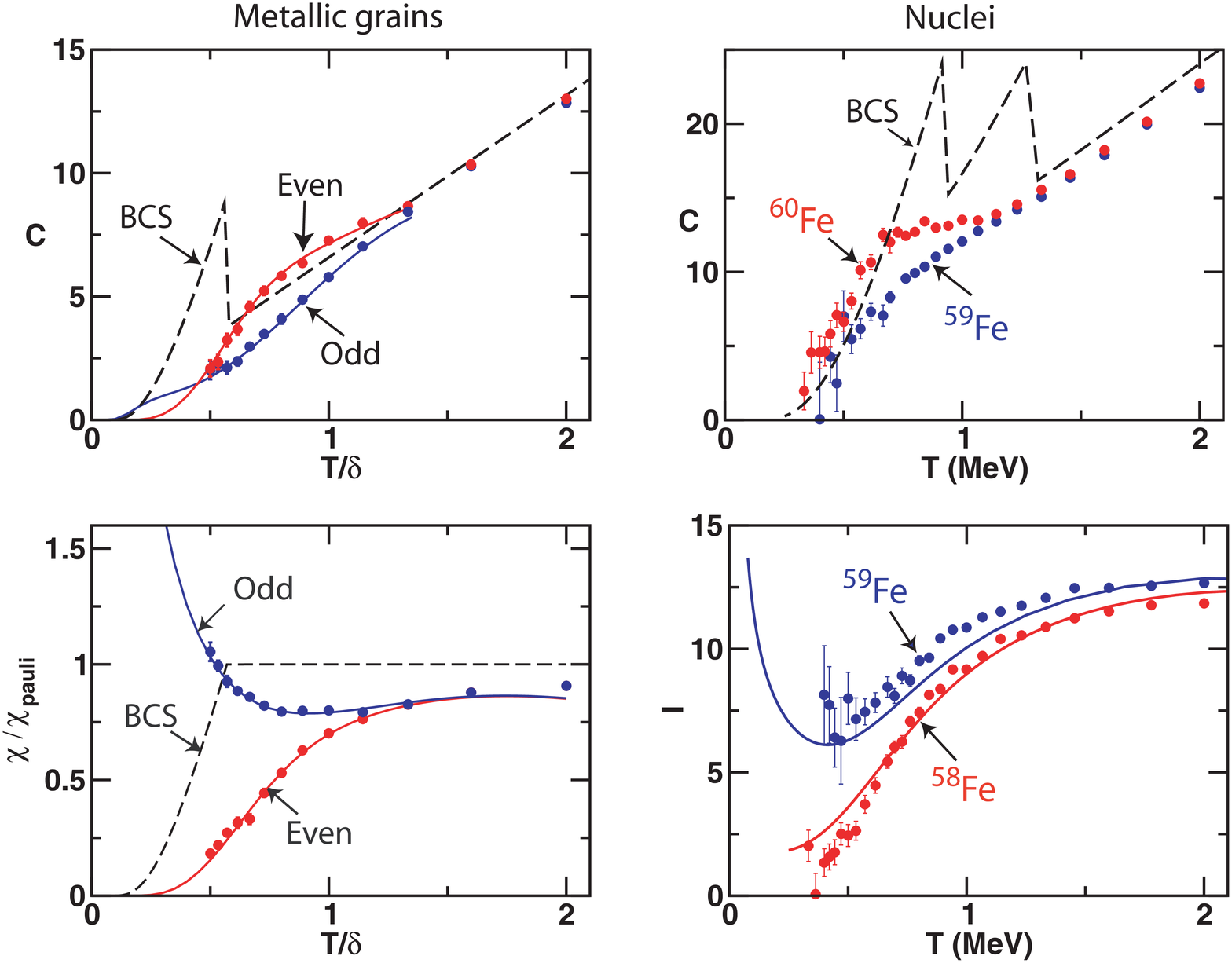,width=12 cm}}
\caption{Thermal signatures of pairing correlations in nano-scale metallic grains with equally spaced single-particle spectrum and $\Delta/\delta=1$ (left panels), and in iron nuclei (right panels). Top panels: the heat capacity. Bottom panels: the spin susceptibility (bottom left for a metallic grain) and the thermal moment of inertia (bottom right for iron nuclei).  The symbols are AFMC calculations. The solid lines in the right panels (metallic grains) are calculated from Richardson's solution. The solid lines in the bottom right panel (nuclei) are from the simple model of Ref.~\citeonline{Alhassid05}. The dashed lines are the results of the BCS approximation.}
\label{nuclei-nanoparticles}
\end{figure}

We discussed thermal signatures of pairing correlation in nuclei and nano-scale metallic grains in the crossover between the bulk BCS limit and the fluctuation-dominated regime where BCS theory breaks down. We used methods that go beyond the mean-field approximation: AFMC and a finite-temperature method that takes into account thermal and small-amplitude quantal fluctuations together with spin and number-parity projections. A summary of some of our results is shown in Fig.~\ref{nuclei-nanoparticles}, in which we compare thermodynamic properties of a metallic grain with $\Delta/\delta=1$ to similar thermal properties in iron nuclei.  An odd-even effect in particle number is observed in the heat capacity of both the metallic grain and the nucleus. Similarly, we compare the spin susceptibility (response to an external magnetic Zeeman field) of metallic grains with the moment of inertia (response to rotations) of nuclei. In both metallic grains and nuclei,  we observe a reentrant effect for an odd number of particles.

The results in Fig.~\ref{nuclei-nanoparticles} also demonstrate the large discrepancy between the BCS approximation and the exact results for various thermodynamic observables, emphasizing the necessity to use methods beyond the mean-field approximation.

\section*{Acknowledgments}
This work was supported in part by the Department of Energy grant DE-FG-0291-ER-40608. I  would like to thank G.F. Bertsch, L. Fang, S. Liu, H. Nakada, K. Nesterov and S. Schmidt for their collaboration on the work presented above.


\begin{thebibliography}{99}
\bibitem{BCS} J. Bardeen, L.N. Cooper and J.R. Schrieffer,
Phys. Rev. {\bf 108},  1175  (1957).
\bibitem{BMP58} A. Bohr, B.R. Mottelson and D. Pines, Phys. Rev. {\bf 110}, 936 (1958).
\bibitem{Belayev59} S.T. Belayev, Mat. Fys. Medd. Dan. Vid. Selsk. {\bf 31} (No. 11) (1959).
\bibitem{nano-experiments} D.C. Ralph, C.T. Black, and M. Tinkham, Phys. Rev. Lett
{\bf 74}, 3241 (1995); {\em ibid.} {\bf 76}, 688 (1996). {\em ibid.} {\bf 78}, 4087 (1997).
\bibitem{vondelft01} J.von Delft and D.C. Ralph, Phys. Rep. {\bf 345},  661 (2001).
\bibitem{Kuemmeth2008}  F.~Kuemmeth, K.~I. Bolotin, S.~F. Shi, and D.~C. Ralph,  Nano Lett. {\bf 8}, 4506 (2008).
\bibitem{Anderson59} P. W. Anderson, J. Phys. Chem. Solids {\bf 11}, 26 (1959).
\bibitem{DiLorenzo2000} A. Di Lorenzo, R. Fazio, F. W. J. Hekking, G. Falci, A. Mastellone, and G. Giaquinta, Phys. Rev. Lett. {\bf 84}, 550 (2000).
\bibitem{Falci2000} G. Falci, R. Fazio, F. W. J. Hekking, and A. Mastellone, J. Low Temp. Phys. {\bf 118}, 355 (2000).
\bibitem{HS} J. Hubbard, Phys. Rev.Lett. {\bf 3}, 77 (1959); R.L.
Stratonovich, Dokl. Akad. Nauk. S.S.S.R. {\bf 115}, 1097 (1957).
\bibitem{Muhlschlegel1972} B. M\"uhlschlegel, D. J. Scalapino, and R. Denton, Phys. Rev. B {\bf 6}, 1767 (1972).
\bibitem{SPA} Y. Alhassid and J. Zingman, Phys Rev. C {\bf 30}, 684 (1984);
 B. Lauritzen, P. Arve, and G.F. Bertsch, Phys. Rev. Lett. {\bf 61}, 2835 (1988); Y. Alhassid and B. Bush,  Nucl. Phys. A {\bf 565},  399 (1993).
\bibitem{GDR} Y. Alhassid, in {\it New Trends
 in Nuclear Collective Dynamics}, Y. Abe  ed.,  Springer Verlag,  NY (1992); Y. Alhassid, Nucl. Phys. A {\bf 649}, 107c (1999),
and references therein.
\bibitem{Kerman1981} A. K. Kerman and S. Levit, Phys. Rev. C \textbf{24}, 1029 (1981); A. K. Kerman, S. Levit, and T. Troudet, Ann. Phys. (NY) \textbf{148}, 436 (1983).
\bibitem{Lauritzen1990} B. Lauritzen, G. Puddu, P. F. Bortignon, and R. A. Broglia, Phys. Lett. B {\bf 246}, 329 (1990)
\bibitem{Puddu1991} G. Puddu, P. Bortignon, and R. Broglia, Ann. Phys. (NY) {\bf 206}, 409 (1991).
\bibitem{Lauritzen1993} B. Lauritzen, A. Anselmino, P. Bortignon, and R. Broglia, Ann. Phys. (NY), {\bf 223}, 216 (1993).
\bibitem{Rossignoli1997} R. Rossignoli and N. Canosa, Phys. Lett. B {\bf 394}, 242 (1997).
\bibitem{Attias1997} H. Attias and Y. Alhassid, Nucl. Phys. A {\bf 625}, 565 (1997).
\bibitem{LG92} E. Y. Loh, Jr. and J. E. Gubernatis, in {\it Electronic
Phase Transitions}, edited by W. Hanke and Y. V. Kopaev (North
Holland, Amsterdam, 1992).
\bibitem{Lang93} G.H. Lang, C.W. Johnson, S.E. Koonin, and W.E. Ormand,
Phys. Rev. C {\bf 48},  1518 (1993).
\bibitem{Alhassid94} Y. Alhassid, D.J. Dean, S.E. Koonin, G.H. Lang, and W.E. Ormand, Phys. Rev. Lett. {\bf 72}, 613 (1994).
\bibitem{Koonin97} S.E. Koonin, D.J. Dean, and K. Langanke, Phys. Rep.
{\bf 278}, 1 (1997).
\bibitem{Alhassid01} Y. Alhassid, Int. J. Mod. Phys. B {\bf 15}, 1447 (2001).
\bibitem{BM69} A. Bohr and B. R. Mottelson, {\it Nuclear Structure},
vol. 1 (Benjamin, New York, 1969).
\bibitem{zuker}  M. Dufour and  A.P. Zuker, Phys. Rev. C {\bf  54}, 1641 (1996).
\bibitem{NA97} H. Nakada and Y. Alhassid, Phys. Rev. Lett. {\bf 79},  2939 (1997).
\bibitem{ABDK96} Y. Alhassid, G.F. Bertsch, D.J. Dean and S.E. Koonin,
Phys. Rev. Lett.  {\bf 77},  1444 (1996).
\bibitem{la01} S. Liu and Y. Alhassid, Phys. Rev. Lett. {\bf 87}, 022501  (2001).
\bibitem{Alhassid03} Y. Alhassid, G.F. Bertsch, and L. Fang, Phys. Rev. C {\bf 68}, 044322 (2003).
\bibitem{schiller01}  A. Schiller, A. Bjerve, M. Guttormsen, M. Hjorth-Jensen, F. Ingebretsen,  E. Melby, S. Messelt, J. Rekstad, S. Siem, and S.W. Odegard, Phys. Rev. C {\bf 63}, 021306  (2001).
 \bibitem{Alhassid07} Y. Alhassid, S. Liu and H. Nakada, Phys. Rev. Lett. {\bf 99}, 162504 (2007).
\bibitem{vonEgidy08} T. von Egidy and D. Bucurescu, Phys. Rev. C {\bf 78},
051301(R) (2008).
\bibitem{er60} T. Ericson, Adv. Phys. {\bf 9}, 425 (1960).
\bibitem{gr74} S.M. Grimes, J.D. Anderson, J.W. McClure, B.A. Pohl and C.
  Wong, Phys. Rev. C {\bf 10}, 2373 (1974).
\bibitem{Alhassid05} Y. Alhassid, G.F. Bertsch, L. Fang and S. Liu, Phys. Rev. C {\bf 72}, 064326 (2005).
\bibitem{Alhassid2000} Y. Alhassid, Rev. Mod. Phys. {\bf 72}, 895 (2000).
\bibitem{Kurland2000} I. L. Kurland, I. L. Aleiner, and B. L. Altshuler, Phys. Rev. B {\bf 62}, 14886 (2000).
\bibitem{Aleiner2002} I. L. Aleiner, P. W. Brouwer, and L. I. Glazman, Phys. Rep. {\bf 358}, 309 (2002).
\bibitem{Alhassid2005} Y. Alhassid, H. A. Weidenm\"uller, and A. Wobst, Phys. Rev. B {\bf 72}, 045318 (2005).
\bibitem{Ying2006} Z. Ying, M. Couco, C. Noce, and H. Zhou, Phys. Rev. B {\bf 74}, 012503 (2006).
\bibitem{Schmidt2007} S. Schmidt, Y. Alhassid, and K. Van Houcke, Europhys. Lett. {\bf 80}, 47004 (2007).
\bibitem{Falci2002} G. Falci, A. Fubini, and A. Mastellone, Phys. Rev. B {\bf 65}, 140507 (2002).
\bibitem{VanHoucke2006} K. Van Houcke, S. M. A. Rombouts, and L. Pollet, Phys. Rev. B {\bf 73}, 132509 (2006).
\bibitem{Schechter2001} M. Schechter, Y. Imry, Y. Levinson, and J. von Delft, Phys. Rev. B {\bf 63}, 214518 (2001).
\bibitem{Alhassid2007} Y. Alhassid, L. Fang, and S. Schmidt, cond-mat/0702304.
\bibitem{Berger1998} S. D. Berger and B. I. Halperin, Phys. Rev. B \textbf{58}, 5213 (1998).
\bibitem{VanHoucke2010} K. Van Houcke, Y. Alhassid, S. Schmidt, and S. Rombouts, arXiv:1011.5421.
\bibitem{richardson} R.W. Richardson, Phys. Rev. Lett. {\bf 3}, 277 (1963); Phys. Rev. {\bf 159}, 792 (1967).
\bibitem{Alhassid2003} Y. Alhassid and T. Rupp, Phys. Rev. Lett. \textbf{91}, 056801 (2003).
\bibitem{Nesterov2012} K. Nesterov and Y. Alhassid, arXiv:1204.5738.
\bibitem{Goodman1981} A.~L. Goodman, Nucl. Phys. A {\bf 352}, 30 (1981).
\bibitem{Rossignoli1998} R.~ Rossignoli, N. Canosa,  and P. Ring, Phys. Rev. Lett. {\bf 80}, 1853 (1998).
\bibitem{Balian1999}  R.~Balian, H. Flocard, and M. Veneroni, Phys. Rep.
  {\bf 317}, 252 (1999).
\bibitem{Ribeiro2012} P. Ribeiro, and A. M. Garc\'ia-Garc\'ia,  Phys. Rev. Lett. {\bf 108}, 097004 (2012).

\end{thebibliography}

\end{document}